\def\lsim{\mathrel{\vcenter{\hbox{$<$}\nointerlineskip\hbox{$\sim$}}}}
\def\gsim{\mathrel{\vcenter{\hbox{$>$}\nointerlineskip\hbox{$\sim$}}}}

\documentclass{elsart}
\usepackage[dvips]{graphicx}
\usepackage{amssymb}
\usepackage{axodraw}
\usepackage{epsf}
\begin{document}
\baselineskip 0.6cm
\newcommand{\be}{\begin{equation}}
\newcommand{\ee}{\end{equation}}
\newcommand{\bea}{\begin{eqnarray}}
\newcommand{\eea}{\end{eqnarray}}

\newcommand{\np}[1]{Nucl. Phys. {\bf #1}}
\newcommand{\pl}[1]{Phys. Lett. {\bf #1}}
\newcommand{\pr}[1]{Phys. Rev. {\bf #1}}
\newcommand{\prl}[1]{Phys. Rev. Lett. {\bf #1}}
\newcommand{\zp}[1]{Z. Phys. {\bf #1}}
\newcommand{\prep}[1]{Phys. Rep. {\bf #1}}
\newcommand{\rmp}[1]{Rev. Mod. Phys. {\bf #1}}    
\newcommand{\ijmp}[1]{Int. Jour. Mod. Phys. {\bf #1}}
\newcommand{\mpl}[1]{Mod. Phys. Lett. {\bf #1}} 
\newcommand{\ptp}[1]{Prog. Theor. Phys. {\bf #1}} 
\newcommand{\arns}[1]{Ann. Rev. Nucl. Sci. {\bf #1}}
\newcommand{\anfis}[1]{An. F\'\i s. {\bf #1}}

\def\lsim{\mathrel{\vcenter{\hbox{$<$}\nointerlineskip\hbox{$\sim$}}}}
\def\gsim{\mathrel{\vcenter{\hbox{$>$}\nointerlineskip\hbox{$\sim$}}}}
\input{FEYNMAN}
\linespread{1.7}

\begin{frontmatter}

\hfill HIP-2000-06/TH

\title{\bf Singly charged higgses at linear collider}

\author{K. Huitu$^{a,1}$,
J. Laitinen$^{a,2}$, 
J.Maalampi$^{b,3}$ and
N. Romanenko$^{b,4}$}

\address{$^a$Helsinki Institute of Physics and 
$^b$Department of Physics}

\address{P.O.Box 9, FIN-00014 University of Helsinki,
Finland}

\thanks[kati]{katri.huitu@helsinki.fi}
\thanks[jouni]{jouni.laitinen@helsinki.fi}
\thanks[jukka]{jukka.maalampi@helsinki.fi}
\thanks[kolya]{nikolai.romanenko@helsinki.fi}

\begin{abstract}
We consider the production of singly charged Higgs bosons
in the Higgs triplet and two Higgs doublet models. 
We evaluate the cross sections for the pair production and the
single production of charged higgses at linear collider. 
The decay modes of $H^+$ and the Standard Model backgrounds are considered. 
We analyze the possibilities to differentiate between triplet and two
Higgs doublet models. 
\bigskip
\end{abstract}


\end{frontmatter}


\section{Introduction}

The observation of a singly charged Higgs boson in future collider experiments would be a definitive evidence of 
physics beyond the Standard Model (SM). The scalar sector of the SM
consists of  a single isodoublet $\phi=(\phi^+,\phi^0)$, and there is only one physical scalar field in the theory, a neutral Higgs. 
The charged scalar degrees of freedom are eaten by the weak boson $W^{\pm}$ via the Higgs mechanism, and they do not show up in 
the spectrum of physical particles. Nevertheless, a common feature of almost all imaginable extensions of the SM 
is the existence of at least one physical charged Higgs field. 

In this paper we shall study the production of a singly charged Higgs scalar in $e^+e^-$ collisions in a future 
linear collider. We shall concentrate on two particular models of different types that both predict one charged 
Higgs scalar, a model with two $SU(2)_L$ Higgs doublets (2HDM) and a model with an $SU(2)_L$ triplet 
(HTM). 

The best known example of the 2HDM is the minimal supersymmetric extension of  the Standard Model (MSSM), which 
requires the existence of two isodoublets of Higgs fields to give masses separately to up and down-type fermions 
and to cancel anomalies  \cite{MSSM}. We will use it 
as our framework when analysing the two-doublet model and implement the
mass relations it predicts in calculating the branching ratios of
charged Higgs decays. We will assume, however,  that the supersymmetric partners of SM particles are too
heavy to affect the processes we will consider.

The triplet model we will consider is a simple extension of the SM, with a Higgs sector including, in addition to the ordinary isodoublet,  a complex 
$\Delta =
(\delta^{++},\delta^+,\delta^0)$ \cite{Godbole} isotriplet with hypercharge $Y=2$ . This kind of a model is 
particularly interesting for it can create  mass to neutrinos via spontaneous symmetry breaking due to Yukawa 
coupling of the triplet with leptons, without requiring the existence of sterile right-handed neutrinos. 
(A strong evidence of non-vanishing neutrino masses was recently obtained in the Super-K measurements of the 
atmospheric neutrino fluxes \cite{superK}.) The vacuum expectation value of  the $\delta^0$ field breaks the
 global lepton number symmetry and
  could in principle lead to the existence of a majoron,
   as in the case 
 of the so-called triplet majoron model \cite{majoron}. The triplet majoron is, however, ruled out experimentally 
 by the LEP measurements of the invisible width of the Z boson
\cite{LEP}. 
 We will consider  the version of the
 HTM where  the
 lepton number is explicitly broken
and no majoron  appears \cite{Godbole}.
This model serves as an example of  the class 
of models which do not contain a majoron.

Isotriplet Higgs scalars arise naturally in the left-right symmetric model \cite{lrm}. 
Although this model differs from the triplet
 model we will consider due to the extended gauge symmetry, 
the phenomenology of the singly charged Higgs particle is quite similar in both models. One feature common to all 
triplet models is that the charged triplet Higgs does not couple to quarks, in contrast with the charged
higgses 
in isodoublets. The physical charged higgses are,
 however, mixtures of isotriplet and isodoublet
higgses, the 
magnitude of the mixing depending on the ratio of the vacuum expectation values of the corresponding neutral Higgs 
fields, so that they do couple to quarks
 but  with a much smaller strength
  than  the ordinary higgses.

The object of our study are the phenomenological features
that would distinguish 
between  purely isodoublet
 and predominantly isotriplet singly charged higgses
   in $e^+e^-$ collisions. 
The possibility 
of the their detection
  has been previously studied in
\cite{Godbole,cheung},  where the effects of  the tree-level
$H^\pm W^\mp Z$ vertex, present in triplet models
but absent in doublet
models, were considered.
This vertex is suppressed in the model with one
complex $Y=2$ triplet \cite{Godbole} considered in this paper, but may
be sizeable in a model which has in addition a real $Y=0$ triplet
\cite{yrjo}.
In the previous studies the possible couplings of triplet charged Higgs
to leptons were not considered (the model in \cite{yrjo} with
real and complex triplets does not have tree-level couplings of
charged Higgs with fermions).
We study the effects of the non-vanishing triplet Yukawa couplings,
which are essential for the neutrino mass generation.

We have organized this paper as follows. In Section 2
 we briefly describe 2HDM and HTM  models.
In Section 3 we investigate the decay of
 the singly charged higgses in
both models. 
 The production
 processes, as well as the SM background,  
 are studied  in Section 4.
  Section 5 is devoted to the summary and conclusions.

\section{The models}

In this chapter we will define
 and summarize several basic features of 
  the two above mentioned  models.
   We will also discuss
   the existing experimental bounds
   on the parameters of these models.

\medskip

{\it The two Higgs doublet model (2HDM).}
The conventional framework to study the singly charged higgses has been the Standard Model with an extended scalar sector consisting of two Higgs isodoublets. 
Usually a discrete symmetry between the doublet Higgs fields $\phi_1$
and $\phi_2$ has been invoked 
in order to avoid the flavour changing
neutral currents \cite{MSSM}.
Depending on the symmetry, the resulting models have been called 
Model I and Model II \cite{MSSM}.
The Higgs sector in Model II includes that of the Minimal 
Supersymmetric Standard Model (MSSM).
In our calculations of the 2HDM, we will use the Model II.

The two Higgs doublets 
\be
\phi_1=\left(\begin{array}{c} \phi_1^+ \\ \phi_1^0\end{array}\right),\;\;\; 
\phi_2=\left(\begin{array}{c}\phi_2^0 \\ \phi_2^-\end{array}\right),
\ee 
have hypercharges $Y=1$ and $-1$, respectively.
The VEV of $\phi_1$ gives masses to down-type quarks and leptons and
$\phi_2$ gives masses to up-type quarks. 
A combination of the charged components of the doublets, $\phi_1^\pm$ and $\phi_2^\pm$, corresponds to the Goldstone
modes $G^\pm$ needed to have $W^\pm$ massive. The other combination orthogonal to this 
forms a physical charged Higgs $H^\pm$. Its couplings to leptons are given by
\be \label{coupl}
\frac{m_l g}{2 \sqrt{2} M_W} \tan \beta\, \overline{\nu_l}(1+\gamma_5) l H^+,
\ee
where $\tan \beta$ is defined as 
\be
\tan \beta =\frac{\langle\phi_2^0\rangle}{ \langle\phi_1^0\rangle} =\frac{v_2 }{v_1}.
\ee
Correspondingly, the couplings of the charged Higgs with quarks in the 
2HDM are given by 
\be
{ {L}}=\frac{g}{2 \sqrt{2} M_W} ( V_{ij} m_{Ui} \cot 
\beta\, \overline{u_i}(1-\gamma_5) d_j+
V_{ij} m_{Dj} \tan \beta \,\overline{u_i} (1+ \gamma_5) d_j)H^+
\ee
where $V_{ij}$ are the CKM matrix elements. 

\medskip

{\it The  Higgs triplet model (HTM).}
The Higgs content  of the HTM
 is more complicated than that of the 2HDM.
    The model
   contains at least one
    Higgs triplet with weak 
   hypercharge $Y=2$ which has 
   lepton number violating couplings
    to leptons but  does not couple to quarks. 
In addition to the triplet, the model
 should contain an
 ordinary
 SM doublet to create
  fermion masses.
The minimal Higgs multiplet content of the triplet model is thus \cite{Godbole}
\bea
\Delta=
\left( \begin{array}{cc}
\delta^+/\sqrt{2} & \delta^{++} \\
\delta^0 & -\delta^+/\sqrt{2} 
\end{array} \right), \,\,\, \phi=
\left( \begin{array}{c}
\phi^+ \\
\phi^0 \end{array} \right).
\eea
Here the neutral components of the triplet and doublet multiplets can
get VEV's, denoted as $\langle\delta^0\rangle = w/\sqrt{2}$ and
$\langle\phi^0\rangle =v/\sqrt{2}$.

 In the 2HDM the electroweak $\rho$-parameter is equal to unity
  at the tree-level while
the loop-level contributions  slightly increase
 this value , but within
the experimental limits.
In contrast to this, in the triplet model 
 the  $\rho$-parameter is less than unity
   already at 
the tree-level due to the isotriplet
 contributions to the masses of electroweak bosons.
  This forces the VEV $w$ of the triplet Higgs to be small compared with the VEV $v$ of the doublet.
We will assume $w  \lsim 15$ 
GeV. \footnote{\baselineskip 0.3cm
It corresponds to the 3-$\sigma$ of the
modern LEP bounds \cite{LEP}. Besides this
we would like to note that
in \cite{BH} the value of the $\rho$-parameter was
discussed in a model with
an additional $Y=0$ triplet.
It was found that at one-loop level the $\rho$-parameters in the
SM and the triplet model are experimentally indistuinguishable.}
Our results are, however, not very sensitive to the actual value of $w$.

The physical charged Higgs will be a mixture of the triplet and
doublet charged components,
\be
\left(\begin{array}{c} H^+ \\G^+ \end{array}\right)=
\left(\begin{array}{cc} -\sin\theta_H & \cos\theta_H \\
\cos\theta_H & \sin\theta_H \end{array}\right)
\left(\begin{array}{c} \phi^+ \\\delta^+ \end{array}\right),
\label{sH}\ee
where
\be
\sin\theta_H=\frac{\sqrt{2} w}{\sqrt{v^2+2 w^2}}, 
\quad \cos\theta_H=\frac{v}{\sqrt{v^2+2 w^2}}.
\ee
The gauge interactions of the triplet Higgs are governed by the term 
\be
{{L}}_\delta^{kin}={\rm Tr} \left\{ (D_\mu \Delta)^\dagger (D^\mu \Delta) \right\}
\ee
with 
\be
D_\mu \Delta=\partial_\mu \Delta + i g' B_\mu+\frac{i g}{2} W_\mu^a [\tau^a,\,\Delta].
\ee

A characteristic feature of the models involving Higgs triplets is the existence of 
the $WZH$-coupling at tree-level. 
This is different from the SM or the \mbox{2HDM},
where $WZH$-coupling does not exist at
tree-level because of the isospin invariance \cite{GM}.
The vertex can appear in these models at loop-level, but it has been found to be too
small to be observed in the case of MSSM \cite{MP}.
Nevertheless, in a general 2HDM, the vertex can be enhanced \cite{Kanemura}.
For large $\tan\beta $ and large mass differences between $H^+$ and
$A$ the branching ratio $H^\pm\rightarrow W^\pm Z$ can be
${\rm {O}}(10^{-2})$.
In this work we consider the parameter region which is consistent with 
the MSSM.
Obviously, the coupling $\gamma W^+ H^-$ is absent in both models at tree-level because 
of the electromagnetic gauge invariance. 

The lepton-triplet Yukawa couplings are given by 
\be \label{eq1}
{{L}}=-i h_{ll'} \Psi_{l L}^T C \tau_2 \Delta\Psi_{l' L}+{ h.c.}, 
\ee
where $C$ is the charge conjugation matrix
and $ \Psi_{l L}$ denotes the left-handed lepton
doublet with flavour $l$. 
After  symmetry breaking, the interaction (\ref{eq1}) yields lepton 
number-violating Majorana masses $m_{\nu_l}=h_{ll} w$
for neutrinos.
The experimental upper bounds for the neutrino masses are \cite{numass}
$m_{\nu_e}\lsim 2.3$ eV, $m_{\nu_\mu}\lsim 170$ keV and 
$m_{\nu_\tau}\lsim 18.2$ MeV.
The values of $h_{ll}$ and $w$
which  we will use in our calculations are consistent with these limits.

For the first and second generation there exist experimental
constraints on the Yukawa couplings $h_{ll'}$ \cite{SLPM}.
Stringent constraints come from the non-observation of the decays $\mu\rightarrow \bar{e} ee$ and
$\mu\rightarrow e\gamma$:
\bea
h_{e\mu}h_{ee}&<&3.2\times 10^{-11}\,{\rm GeV}^{-2} M_{\delta^{++}}^2,
\nonumber\\
h_{e\mu}h_{\mu\mu}&<&2\times 10^{-10}\,{\rm GeV}^{-2} M_{\delta^{++}}^2.
\eea
{}From Bhabha scattering follows an upper limit for $h_{ee}$,
\be
h_{ee}^2\lsim 9.7\times 10^{-6}\,{\rm GeV}^{-2} M_{\delta^{++}}^2,
\label{hee}\ee
and from $(g-2)_\mu$ an upper limit for $h_{\mu\mu}$,
\be
h_{\mu\mu}^2\lsim 2.5\times 10^{-5}\,{\rm GeV}^{-2} M_{\delta^{++}}^2.
\label{hmm}\ee
{}From muonium-antimuonium transition one finds the following
bound:
\be
h_{ee}h_{\mu\mu}\lsim 5.8\times 10^{-5}\,{\rm GeV}^{-2} M_{\delta^{++}}^2.
\label{hem}\ee
For the third generation Yukawa couplings $h_{\tau\tau}$,
$h_{\tau e}$ and $h_{\tau\mu}$ there are no limits so far.

We will assume that the Yukawa couplings $h_{ll'}$
 are diagonal. One deduces
  from the above constraints that
   all values of the diagonal coupling
    constants below the perturbation limit
 of $h_{ll}\lsim 1$  are allowed if the mass of the doubly charged Higgs exceeds 300 GeV.

The interactions of $H^+$ with the leptons are given by 
\be \label{inthenu} 
\frac 12 \overline{\nu}_l \left[ h_{ll} \cos{\theta_H} (1-\gamma_5)-\sin{\theta_H}
\frac{\sqrt{2} m_l}{v} (1+\gamma_5) \right] l H^{+}+h.c.
\ee  
The latter term, which originates from the Yukawa couplings of the 
isodoublet, is suppressed by both the small value of $\sin{\theta_H}$ 
and the ratio $m_l/v$. 
It will be neglected in the following.

The triplet Higgs $\delta^+$ has no couplings to  quarks due to conservation of weak hypercharge.
Nevertheless, the physical charged Higgs $H^+$ does have  couplings to 
quarks because of the doublet-triplet mixing (\ref{sH}), but they are suppressed by the small
mixing angle $\sin\theta_H$.

The classical scalar potential of the model (see Eq. (2) of ref. \cite{Godbole}) contains
among other terms a dimension-3 coupling:
\be
\phi^T \tau_2 \Delta^\dagger
 \phi+ { h.c.},
\ee

which breaks lepton number explicitly. 
As discussed in the introduction, this term is necessary
to avoid the existence of  majorons. 
The Higgs potential implies the following relations of various scalar masses:
\begin{eqnarray}
M_{\delta^{++}}^2=M_{H^+}^2-\frac{\lambda_5}{2}v^2 \\
\frac{1}{2}M_{H^0_{im}}^2=M_{H^+}^2+\frac{\lambda_5}{2}v^2 \nonumber \\
\end{eqnarray}
where $H^0_{im}$ is the physical combination of Im($\delta^0$) and Im($\phi^0$) and $\lambda_5$ is a coupling constant appearing in the scalar potential. One can conclude from these relations that the doubly charged and singly charged
higgses are quite degenerate in mass, except when the coupling constant $\lambda_5$ happens to be  large.

\section{Decay modes of $H^\pm$}

\begin{figure}[t]
\leavevmode
\includegraphics[width=14cm, height=10cm]{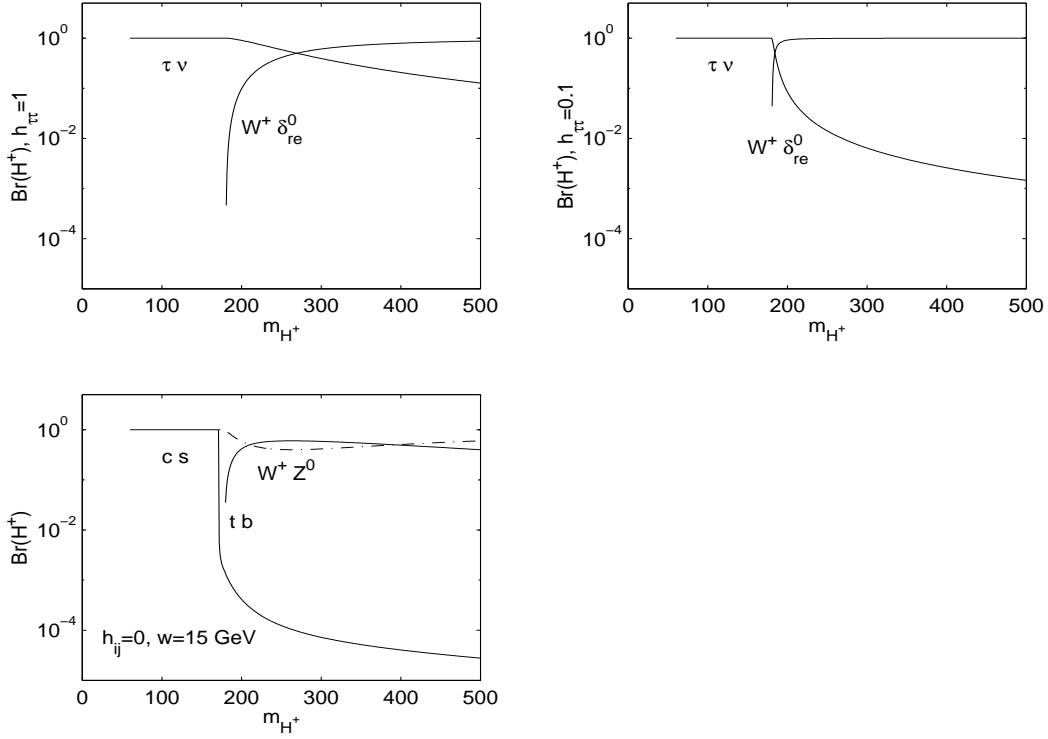}
\caption{\label{decays} Brancing ratios of $H^+$
 in the HTM, with $h_{\tau \tau}= 1, 0.1$ and $0$.}
\end{figure} 

\indent
Signatures of  charged higgses  produced are determined by their decay modes. Depending on 
the unknown parameters, like scalar masses, 
$\tan\beta$ and the triplet Higgs Yukawa
 couplings $h_{ll'}$, the decay modes 
and their branching ratios may be quite different
 in the 2HDM and HTM. 
 In both models there exist the following two-body decay channels:
\be
H^+ \rightarrow \overline{l} \nu_l,\, u_i \overline{d}_i,\, W^+H^0, \, W^+ Z^0. 
\ee
The branching ratios in the HTM are presented
 in Fig.1 for two values of
  the third generation Majorana Yukawa
   coupling ($h_{\tau\tau}= 1$ and 0.1).
    We remind that this Yukawa coupling is not restricted by the present experiments. We have assumed here that the couplings $h_{ee}$ and $h_{\mu \mu}$ are small compared with $h_{\tau\tau}$. Below the boson pair 
threshold the dominant decay channel of the 
charged Higgs is $H^+\rightarrow \tau^+ \nu_{\tau}$.
 The other important channel $H^+\rightarrow c \bar{s}$ 
 is suppressed  by the factor
$\sin{\theta_H}$ as quarks do not couple to the isotriplet component of $H^+$ but only with the isodoublet component.
If the Yukawa coupling $h_{\tau\tau}\lsim 10^{-3}$, the mode $H^+\rightarrow c\overline{s}$ 
would have a comparable or larger branching ratio
 than the mode $H^+\rightarrow \tau^+ \nu_{\tau}$. For the 
values $h_{\tau\tau}= 0.1\; - 1$ the $c\bar s$-channel is negligible. The same is true if one has $h_{ee}$ or $h_{\mu\mu}= 0.1\; - 1$, which are consistent with the bounds (\ref{hee}), (\ref{hmm}) and  (\ref{hem})  when $m_{H^{\pm}}\gsim 300$ GeV.\\ 

\indent
When the mass of $H^+$ is above the  $H^0 W$ threshold (where $H^0$ is either the pseudoscalar neutral Higgs 
or the real part of the neutral triplet $\delta^0$), the decay mode $H^+ \rightarrow W^+ H^0$ quickly takes over  the $\tau\nu_{\tau}$-channel
and becomes the dominant channel.
The $W$ boson produced  decays into a pair of quarks or a lepton and a neutrino, 
and the neutral Higgs, if it is  $\delta^0_{re}$, decays most of the time  invisibly into neutrinos. 
The signal of this decay mode will 
therefore be a charged lepton plus missing energy,
 just like in the direct leptonic decay $H^+ \rightarrow
 l^+ \nu_l$, 
or a pair of
 quarks with invariant mass $M_W$ plus missing energy.  
There exists also the decay channel
 to $W^+ \phi^0_{re}$, where $\phi^0_{re}$ is the real part 
of the neutral doublet Higgs (this is the one which resembles the SM Higgs), but  
it is less important because  the relevant vertex is suppressed by the factor
$\sin\theta_{H}$. (It is ignored in the Fig.1.) 
The interactions $H^+ W^- Z^0$  and $H^+q \overline{q^´}$
 are also suppressed by the doublet-triplet mixing angle 
$\sin\theta_H$, and the corresponding branching ratios are therefore too small to be visible in  Fig. 1.\\

\indent
We depict in Fig. 1  the branching ratios also when all
the Majorana Yukawa couplings $h_{ll'}$ are set to zero. This changes the decay pattern dramatically for small values of $M_{H^+}$. 
In this case the mode $\tau \nu$ is absent and the channel to $c\bar s$ is the dominant one. For  a heavier $H^+$ also the decays to $W^+Z^0$ and $t\bar b$ can have significant branching ratios depending on the value of  $M_{\delta_0^{re}}$. The channel $H^+ \rightarrow \delta_0^{re}W^+$ will also in this case be the dominant  one, whenever it is kinematically allowed. The branching ratio of the relatively rare decay $W^+H^0$ is not depicted in Fig. 1 for this case of  vanishing Majorana Yukawa couplings. \\

\indent
In the case of  vanishing Majorana Yukawa couplings,  the total width of $H^\pm$ is below the $W\delta_0^{re}$ threshold
 proportional to $\sin^2 \theta_H$ . Consequently, if the doublet-triplet mixing is small enough, $H^+$ would leave the detector before it decays. This kind of semi-stable charged particle would give a clear signature. \\

\indent 
The branching ratios for the case of 2HDM,
 in the framework of the \mbox{MSSM},  have been extensively studied e.g. in ref. \cite{Ritva}
It is found that for  a low $M_{H^+}$, the decays to $c\bar s$ or $\tau \nu_\tau$ are the most important ones.  
Also the decays $H^+ \rightarrow A^0 W^+,\,h^0 W^+$, with virtual Higgs bosons, may have sizeable branching ratios 
in the case of a light $H^+$ and a small $\tan \beta$.  
The channel $H^+\to t\overline{b}$ is always the dominant channel when $M_{H^+}$ exceeds the appropriate 
threshold. Actually this channel, with a virtual $t$, may have a substantial branching ratio already below the threshold,
 in particular in the case of small $\tan\beta$. The subdominant mode is $H^+\rightarrow \tau\nu_{\tau}$, but its 
branching ratio is much smaller, in particular for small values of $\tan\beta$, and the $H^+$ decays almost 
exclusively to the $t\bar b$ pair. In the presence of  SUSY, new decay modes may open, 
providing that there exist light enough SUSY particles. For example, the two doublet case is realized in a SUSY model, 
e.g. in the MSSM, the decays of $H^+$ into lightest neutralinos and charginos might be kinematically allowed for large 
values of $M_{H^+}$. This would drastically suppress the branching ratio of the $H^+\to tb$ channel \cite{Borzu}. \\

\indent 
To summarise, in the large mass range and if at least one of the Yukawa couplings is not extremely small, 
a charged Higgs of the HTM and that of the 2HDM can be easily distinguished 
from each other through their decay characteristics, the final state being  $ l^+$ + missing energy  in the former 
case and $t\overline{b}$ in the latter case. In the low mass range the situation is less clear, the dominant channel 
in both cases being to $\tau\nu_{\tau}$, if we assume that $h_{\tau \tau}$ is the largest Majorana Yukawa coupling. The absence of 
both the $\tau\nu_{\tau}$ and $c\overline s$ decay modes would be a favourable indication towards the HTM and would imply large $h_{ee}$ or $h_{\mu\mu}$ couplings. In the two doublet case there exists  a sizeable $c\overline s$ component, which the HTM in general misses 
if at least one Majorana Yukawa coupling is of the order 0.1-1.

\section{Production of  $H^\pm$}

Our main emphasis in the following will be placed on the single production of the charged Higgs, but let us start with a comment concerning the pair production $e^+e^- \rightarrow H^+ H^-$.

\begin{figure}[t]
\begin{center}
\vspace{1cm}
\includegraphics[width=14cm, height=15cm]{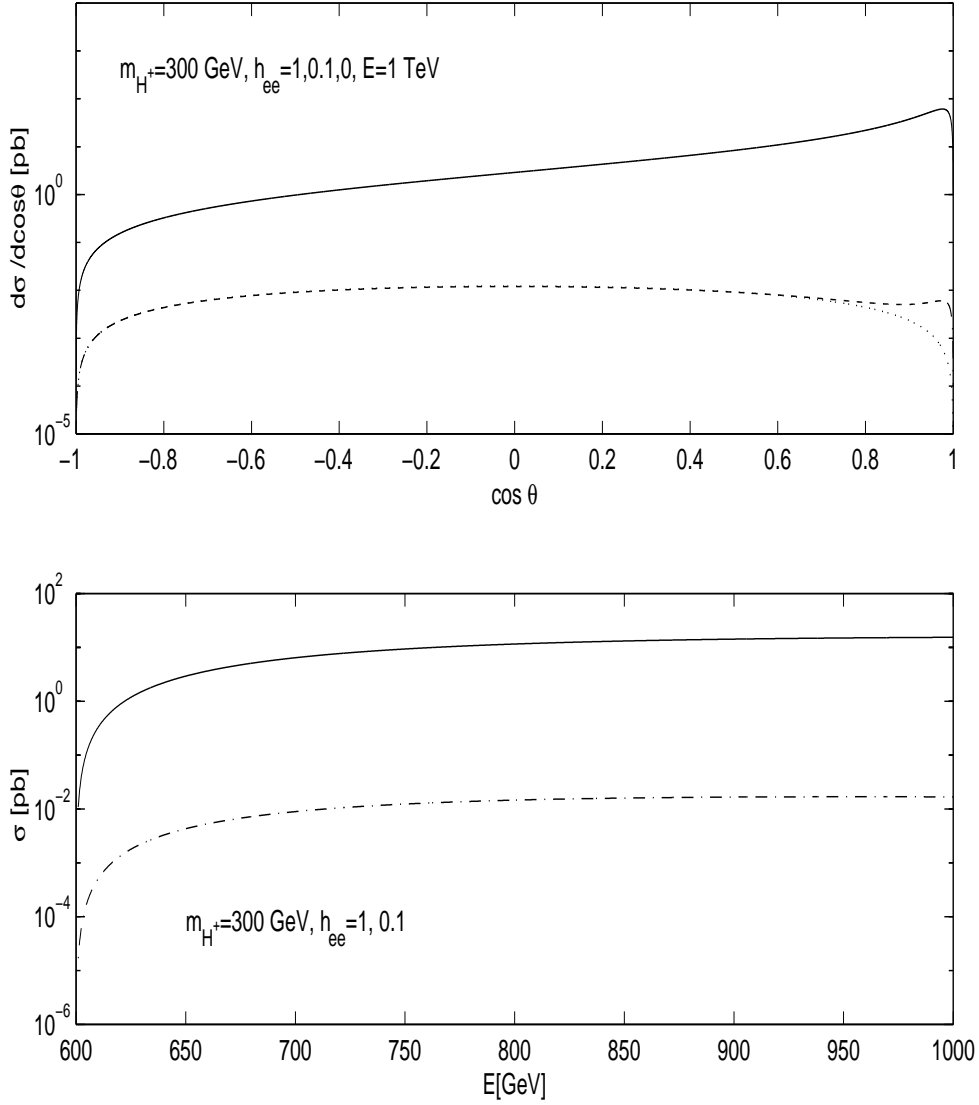}
\caption{\label{pair1}
Angular distribution
for $\sqrt{s}$, $h_{ee}=1$ (solid) $0.1$ (dashed)
$0$ (dotted)
 and energy dependence of the
 cross section $h_{ee}=1$ (solid) $0.1$ (dot-dashed)
in HTM, $m_{H^+}=300$ GeV.} 
\end{center}
\end{figure}

In the 2HDM the pair production occurs through
 a $\gamma$ and $Z^0$ exchange in s-channel.
 The unpolarized cross section is given  e.g. in ref.  \cite{komamiya88}. 
%
%
In the HTM there is an additional amplitude, 
the t-channel neutrino exchange.
In  Fig. \ref{pair1} we present
 the differential cross sections of the process in the HTM for the parameter values $M_H=300$ GeV and $h_{ee}=0,\;0.1$ and 1. The t-channel neutrino 
diagram  increases the pair production cross section by several orders of magnitude if the Yukawa 
coupling $h_{ee}$ has a value close to 1.
 It also changes   the angular distribution from
   the $ \sin^2 \theta$ form obtained in the 2HDM 
    to a forward peaked one.  
As follows from the structure
 of the coupling (\ref{inthenu}),
  the $t$-channel diagram can be
   turned off
   if the electron beam has right-handed polarization.
    This provides
     a method to test
      the existence and the
       magnitude of the neutrino exchange amplitude. 

Let us proceed to the single production of $H^+$. Of course, this production channel opens below the pair production threshold and could therefore be the first place to discover a charged Higgs particle. A two-body associate production occurs through the channel
\be
e^+e^- \rightarrow H^+ W^-.
\ee
This process is mediated 
by the t-channel neutrino exchange and s-channel $Z$-exchange. 
In the triplet model the cross section for this process is,  however, very 
small. This is so because the neutrino-exchange diagram is proportional to the 
electron neutrino mass, which is constrained to be less than a few eV's. 
The cross section remains very small
 for $\sqrt{s}=0.5-2 $ TeV. 
The $Z$-exchange diagram, in turn, has the $\sin\theta_H$-mixing angle
suppression from the $WZH$-vertex.

On the other hand, if the triplet Yukawa coupling is nonvanishing, 
one may expect larger cross sections for the leptonic three-body reactions 
\be \label{process}
e^+e^- \rightarrow l^- \nu_l H^+,\; l=e,\mu,\tau.
\ee

We have evaluated the cross sections for the processes (\ref{process})
in the HTM and 2HDM. 
The relevant Feynman rules for both models were generated with the LanHEP 
package \cite{lanhep} (the Feynman rules for the 2HDM can also be
found in \cite{MSSM}). 
The symbolic and numerical calculations of the cross sections were
carried out using the CompHEP package \cite{comphep}. 

In the 2HDM, the value of  the cross section depends on two unknown 
independent parameters, $M_{H^\pm}$ and $\tan \beta$. 
In the HTM, the unknown model parameters are the three Majorana Yukawa 
couplings $h_{ll}, l=e,\mu,\tau$, the VEV $w$ of the neutral triplet scalar
and the masses of the Higgs bosons. 
For definiteness, the masses of all the neutral and the doubly 
charged scalars are taken to be 300 GeV. 

\begin{figure}[t]
{
\unitlength=0.8 pt
\SetScale{0.8}
\SetWidth{0.56}      
\scriptsize    
{} \qquad\allowbreak
\noindent
\begin{picture}(95,69)(0,0)
\Text(15.0,70.0)[r]{$e^-$}
\ArrowLine(16.0,70.0)(37.0,60.0) 
\Text(15.0,50.0)[r]{$e^+$}
\ArrowLine(37.0,60.0)(16.0,50.0) 
\Text(47.0,61.0)[b]{$A,Z$}
\DashLine(37.0,60.0)(58.0,60.0){3.0} 
\Text(80.0,70.0)[l]{$e^-$}
\ArrowLine(58.0,60.0)(79.0,70.0) 
\Text(54.0,50.0)[r]{$e^-$}
\ArrowLine(58.0,40.0)(58.0,60.0) 
\Text(80.0,50.0)[l]{$\nu_e$}
\Line(58.0,40.0)(79.0,50.0) 
\Text(80.0,30.0)[l]{$H^+$}
\DashArrowLine(58.0,40.0)(79.0,30.0){1.0} 
\end{picture} \ 
{} \qquad\allowbreak
\begin{picture}(95,69)(0,0)
\Text(15.0,60.0)[r]{$e^-$}
\ArrowLine(16.0,60.0)(37.0,60.0) 
\Line(37.0,60.0)(58.0,60.0) 
\Text(80.0,70.0)[l]{$e^-$}
\ArrowLine(58.0,60.0)(79.0,70.0) 
\Text(36.0,50.0)[r]{$A,Z$}
\DashLine(37.0,60.0)(37.0,40.0){3.0} 
\Text(15.0,40.0)[r]{$e^+$}
\ArrowLine(37.0,40.0)(16.0,40.0) 
\Text(47.0,44.0)[b]{$e^-$}
\ArrowLine(58.0,40.0)(37.0,40.0) 
\Text(80.0,50.0)[l]{$\nu_e$}
\Line(58.0,40.0)(79.0,50.0) 
\Text(80.0,30.0)[l]{$H^+$}
\DashArrowLine(58.0,40.0)(79.0,30.0){1.0} 
\end{picture} \ 
{} \qquad\allowbreak
\begin{picture}(95,69)(0,0)
\Text(15.0,70.0)[r]{$e^-$}
\ArrowLine(16.0,70.0)(37.0,60.0) 
\Text(15.0,50.0)[r]{$e^+$}
\ArrowLine(37.0,60.0)(16.0,50.0) 
\Text(47.0,61.0)[b]{$A,Z$}
\DashLine(37.0,60.0)(58.0,60.0){3.0} 
\Text(80.0,70.0)[l]{$H^+$}
\DashArrowLine(58.0,60.0)(79.0,70.0){1.0} 
\Text(54.0,50.0)[r]{$H^+$}
\DashArrowLine(58.0,40.0)(58.0,60.0){1.0} 
\Text(80.0,50.0)[l]{$\nu_e$}
\Line(58.0,40.0)(79.0,50.0) 
\Text(80.0,30.0)[l]{$e^-$}
\ArrowLine(58.0,40.0)(79.0,30.0) 
\end{picture} \ 
{} \qquad\allowbreak
\begin{picture}(95,69)(0,0)
\Text(15.0,70.0)[r]{$e^-$}
\ArrowLine(16.0,70.0)(58.0,70.0) 
\Text(80.0,70.0)[l]{$e^-$}
\ArrowLine(58.0,70.0)(79.0,70.0) 
\Text(57.0,60.0)[r]{$A,Z$}
\DashLine(58.0,70.0)(58.0,50.0){3.0} 
\Text(80.0,50.0)[l]{$H^+$}
\DashArrowLine(58.0,50.0)(79.0,50.0){1.0} 
\Text(54.0,40.0)[r]{$H^+$}
\DashArrowLine(58.0,30.0)(58.0,50.0){1.0} 
\Text(15.0,30.0)[r]{$e^+$}
\ArrowLine(58.0,30.0)(16.0,30.0) 
\Text(80.0,30.0)[l]{$\nu_e$}
\Line(58.0,30.0)(79.0,30.0) 
\end{picture} \ 
{} \qquad\allowbreak
\begin{picture}(95,69)(0,0)
\Text(15.0,70.0)[r]{$e^-$}
\ArrowLine(16.0,70.0)(37.0,60.0) 
\Text(15.0,50.0)[r]{$e^+$}
\ArrowLine(37.0,60.0)(16.0,50.0) 
\Text(47.0,61.0)[b]{$Z$}
\DashLine(37.0,60.0)(58.0,60.0){3.0} 
\Text(80.0,70.0)[l]{$H^+$}
\DashArrowLine(58.0,60.0)(79.0,70.0){1.0} 
\Text(54.0,50.0)[r]{$W^+$}
\DashArrowLine(58.0,40.0)(58.0,60.0){3.0} 
\Text(80.0,50.0)[l]{$\nu_e$}
\Line(58.0,40.0)(79.0,50.0) 
\Text(80.0,30.0)[l]{$e^-$}
\ArrowLine(58.0,40.0)(79.0,30.0) 
\end{picture} \ 
{} \qquad\allowbreak
\begin{picture}(95,69)(0,0)
\Text(15.0,70.0)[r]{$e^-$}
\ArrowLine(16.0,70.0)(58.0,70.0) 
\Text(80.0,70.0)[l]{$e^-$}
\ArrowLine(58.0,70.0)(79.0,70.0) 
\Text(57.0,60.0)[r]{$Z$}
\DashLine(58.0,70.0)(58.0,50.0){3.0} 
\Text(80.0,50.0)[l]{$H^+$}
\DashArrowLine(58.0,50.0)(79.0,50.0){1.0} 
\Text(54.0,40.0)[r]{$W^+$}
\DashArrowLine(58.0,30.0)(58.0,50.0){3.0} 
\Text(15.0,30.0)[r]{$e^+$}
\ArrowLine(58.0,30.0)(16.0,30.0) 
\Text(80.0,30.0)[l]{$\nu_e$}
\Line(58.0,30.0)(79.0,30.0) 
\end{picture} \ 
{} \qquad\allowbreak
\begin{picture}(95,69)(0,0)
\Text(15.0,70.0)[r]{$e^-$}
\ArrowLine(16.0,70.0)(37.0,60.0) 
\Text(15.0,50.0)[r]{$e^+$}
\ArrowLine(37.0,60.0)(16.0,50.0) 
\Text(47.0,61.0)[b]{$Z$}
\DashLine(37.0,60.0)(58.0,60.0){3.0} 
\Text(80.0,70.0)[l]{$\nu_e$}
\Line(58.0,60.0)(79.0,70.0) 
\Text(57.0,50.0)[r]{$\nu_e$}
\Line(58.0,60.0)(58.0,40.0) 
\Text(80.0,50.0)[l]{$H^+$}
\DashArrowLine(58.0,40.0)(79.0,50.0){1.0} 
\Text(80.0,30.0)[l]{$e^-$}
\ArrowLine(58.0,40.0)(79.0,30.0) 
\end{picture} \ 
{} \qquad\allowbreak
\begin{picture}(95,69)(0,0)
\Text(15.0,70.0)[r]{$e^-$}
\ArrowLine(16.0,70.0)(58.0,70.0) 
\Text(80.0,70.0)[l]{$e^-$}
\ArrowLine(58.0,70.0)(79.0,70.0) 
\Text(57.0,60.0)[r]{$Z$}
\DashLine(58.0,70.0)(58.0,50.0){3.0} 
\Text(80.0,50.0)[l]{$\nu_e$}
\Line(58.0,50.0)(79.0,50.0) 
\Text(57.0,40.0)[r]{$\nu_e$}
\Line(58.0,50.0)(58.0,30.0) 
\Text(15.0,30.0)[r]{$e^+$}
\ArrowLine(58.0,30.0)(16.0,30.0) 
\Text(80.0,30.0)[l]{$H^+$}
\DashArrowLine(58.0,30.0)(79.0,30.0){1.0} 
\end{picture} \ 
{} \qquad\allowbreak
\begin{picture}(95,69)(0,0)
\Text(15.0,60.0)[r]{$e^-$}
\ArrowLine(16.0,60.0)(37.0,60.0) 
\Text(47.0,64.0)[b]{$e^-$}
\ArrowLine(58.0,60.0)(37.0,60.0) 
\Text(80.0,70.0)[l]{$\nu_e$}
\Line(58.0,60.0)(79.0,70.0) 
\Text(80.0,50.0)[l]{$H^+$}
\DashArrowLine(58.0,60.0)(79.0,50.0){1.0} 
\Text(33.0,50.0)[r]{$\Delta^{++}$}
\DashArrowLine(37.0,40.0)(37.0,60.0){1.0} 
\Text(15.0,40.0)[r]{$e^+$}
\ArrowLine(37.0,40.0)(16.0,40.0) 
\Line(37.0,40.0)(58.0,40.0) 
\Text(80.0,30.0)[l]{$e^-$}
\ArrowLine(58.0,40.0)(79.0,30.0) 
\end{picture} \ 
{} \qquad\allowbreak
\begin{picture}(95,69)(0,0)
\Text(15.0,60.0)[r]{$e^-$}
\ArrowLine(16.0,60.0)(37.0,60.0) 
\Text(47.0,61.0)[b]{$\nu_e$}
\Line(37.0,60.0)(58.0,60.0) 
\Text(80.0,70.0)[l]{$H^+$}
\DashArrowLine(58.0,60.0)(79.0,70.0){1.0} 
\Text(80.0,50.0)[l]{$e^-$}
\ArrowLine(58.0,60.0)(79.0,50.0) 
\Text(33.0,50.0)[r]{$W^+$}
\DashArrowLine(37.0,40.0)(37.0,60.0){3.0} 
\Text(15.0,40.0)[r]{$e^+$}
\ArrowLine(37.0,40.0)(16.0,40.0) 
\Line(37.0,40.0)(58.0,40.0) 
\Text(80.0,30.0)[l]{$\nu_e$}
\Line(58.0,40.0)(79.0,30.0) 
\end{picture} \ 
{} \qquad\allowbreak
\begin{picture}(95,69)(0,0)
\Text(15.0,70.0)[r]{$e^-$}
\ArrowLine(16.0,70.0)(58.0,70.0) 
\Text(80.0,70.0)[l]{$\nu_e$}
\Line(58.0,70.0)(79.0,70.0) 
\Text(54.0,60.0)[r]{$W^+$}
\DashArrowLine(58.0,50.0)(58.0,70.0){3.0} 
\Text(80.0,50.0)[l]{$H^+$}
\DashArrowLine(58.0,50.0)(79.0,50.0){1.0} 
\Text(54.0,40.0)[r]{$\Delta^{++}$}
\DashArrowLine(58.0,30.0)(58.0,50.0){1.0} 
\Text(15.0,30.0)[r]{$e^+$}
\ArrowLine(58.0,30.0)(16.0,30.0) 
\Text(80.0,30.0)[l]{$e^-$}
\ArrowLine(58.0,30.0)(79.0,30.0) 
\end{picture} \ 
{} \qquad\allowbreak
\begin{picture}(95,69)(0,0)
\Text(15.0,60.0)[r]{$e^-$}
\ArrowLine(16.0,60.0)(37.0,60.0) 
\Line(37.0,60.0)(58.0,60.0) 
\Text(80.0,70.0)[l]{$\nu_e$}
\Line(58.0,60.0)(79.0,70.0) 
\Text(33.0,50.0)[r]{$W^+$}
\DashArrowLine(37.0,40.0)(37.0,60.0){3.0} 
\Text(15.0,40.0)[r]{$e^+$}
\ArrowLine(37.0,40.0)(16.0,40.0) 
\Text(47.0,41.0)[b]{$\nu_e$}
\Line(37.0,40.0)(58.0,40.0) 
\Text(80.0,50.0)[l]{$H^+$}
\DashArrowLine(58.0,40.0)(79.0,50.0){1.0} 
\Text(80.0,30.0)[l]{$e^-$}
\ArrowLine(58.0,40.0)(79.0,30.0) 
\end{picture} \ 
{} \qquad\allowbreak
\begin{picture}(95,69)(0,0)
\Text(15.0,70.0)[r]{$e^-$}
\ArrowLine(16.0,70.0)(58.0,70.0) 
\Text(80.0,70.0)[l]{$\nu_e$}
\Line(58.0,70.0)(79.0,70.0) 
\Text(54.0,60.0)[r]{$W^+$}
\DashArrowLine(58.0,50.0)(58.0,70.0){3.0} 
\Text(80.0,50.0)[l]{$e^-$}
\ArrowLine(58.0,50.0)(79.0,50.0) 
\Text(57.0,40.0)[r]{$\nu_e$}
\Line(58.0,50.0)(58.0,30.0) 
\Text(15.0,30.0)[r]{$e^+$}
\ArrowLine(58.0,30.0)(16.0,30.0) 
\Text(80.0,30.0)[l]{$H^+$}
\DashArrowLine(58.0,30.0)(79.0,30.0){1.0} 
\end{picture} \ 
{} \qquad\allowbreak
\begin{picture}(95,69)(0,0)
\Text(15.0,60.0)[r]{$e^-$}
\ArrowLine(16.0,60.0)(37.0,60.0) 
\Text(47.0,64.0)[b]{$H^+$}
\DashArrowLine(58.0,60.0)(37.0,60.0){1.0} 
\Text(80.0,70.0)[l]{$\nu_e$}
\Line(58.0,60.0)(79.0,70.0) 
\Text(80.0,50.0)[l]{$e^-$}
\ArrowLine(58.0,60.0)(79.0,50.0) 
\Text(36.0,50.0)[r]{$\nu_e$}
\Line(37.0,60.0)(37.0,40.0) 
\Text(15.0,40.0)[r]{$e^+$}
\ArrowLine(37.0,40.0)(16.0,40.0) 
\DashLine(37.0,40.0)(58.0,40.0){1.0}
\Text(80.0,30.0)[l]{$H^+$}
\DashArrowLine(58.0,40.0)(79.0,30.0){1.0} 
\end{picture} \ 
{} \qquad\allowbreak
\begin{picture}(95,69)(0,0)
\Text(15.0,60.0)[r]{$e^-$}
\ArrowLine(16.0,60.0)(37.0,60.0) 
\Text(47.0,61.0)[b]{$\nu_e$}
\Line(37.0,60.0)(58.0,60.0) 
\Text(80.0,70.0)[l]{$H^+$}
\DashArrowLine(58.0,60.0)(79.0,70.0){1.0} 
\Text(80.0,50.0)[l]{$e^-$}
\ArrowLine(58.0,60.0)(79.0,50.0) 
\Text(33.0,50.0)[r]{$H^+$}
\DashArrowLine(37.0,40.0)(37.0,60.0){1.0} 
\Text(15.0,40.0)[r]{$e^+$}
\ArrowLine(37.0,40.0)(16.0,40.0) 
\Line(37.0,40.0)(58.0,40.0) 
\Text(80.0,30.0)[l]{$\nu_e$}
\Line(58.0,40.0)(79.0,30.0) 
\end{picture} \ 
{} \qquad\allowbreak
\begin{picture}(95,69)(0,0)
\Text(15.0,60.0)[r]{$e^-$}
\ArrowLine(16.0,60.0)(37.0,60.0) 
\Line(37.0,60.0)(58.0,60.0) 
\Text(80.0,70.0)[l]{$\nu_e$}
\Line(58.0,60.0)(79.0,70.0) 
\Text(33.0,50.0)[r]{$H^+$}
\DashArrowLine(37.0,40.0)(37.0,60.0){1.0} 
\Text(15.0,40.0)[r]{$e^+$}
\ArrowLine(37.0,40.0)(16.0,40.0) 
\Text(47.0,41.0)[b]{$\nu_e$}
\Line(37.0,40.0)(58.0,40.0) 
\Text(80.0,50.0)[l]{$H^+$}
\DashArrowLine(58.0,40.0)(79.0,50.0){1.0} 
\Text(80.0,30.0)[l]{$e^-$}
\ArrowLine(58.0,40.0)(79.0,30.0) 
\end{picture} \ 
{} \qquad\allowbreak
\begin{picture}(95,69)(0,0)
\Text(15.0,70.0)[r]{$e^-$}
\ArrowLine(16.0,70.0)(58.0,70.0) 
\Text(80.0,70.0)[l]{$\nu_e$}
\Line(58.0,70.0)(79.0,70.0) 
\Text(54.0,60.0)[r]{$H^+$}
\DashArrowLine(58.0,50.0)(58.0,70.0){1.0} 
\Text(80.0,50.0)[l]{$e^-$}
\ArrowLine(58.0,50.0)(79.0,50.0) 
\Text(57.0,40.0)[r]{$\nu_e$}
\Line(58.0,50.0)(58.0,30.0) 
\Text(15.0,30.0)[r]{$e^+$}
\ArrowLine(58.0,30.0)(16.0,30.0) 
\Text(80.0,30.0)[l]{$H^+$}
\DashArrowLine(58.0,30.0)(79.0,30.0){1.0} 
\end{picture} \ 
}
\caption{Feynman diagrams for the process
 $e^+e^- \rightarrow e^- \nu_e H^+$
in the Higgs triplet model.\label{ediag}}
\end{figure}

\begin{figure}[t]
\vspace*{-0.1cm} \hspace*{-0.3cm}
\epsfysize=10cm \epsfxsize=7cm
\epsffile{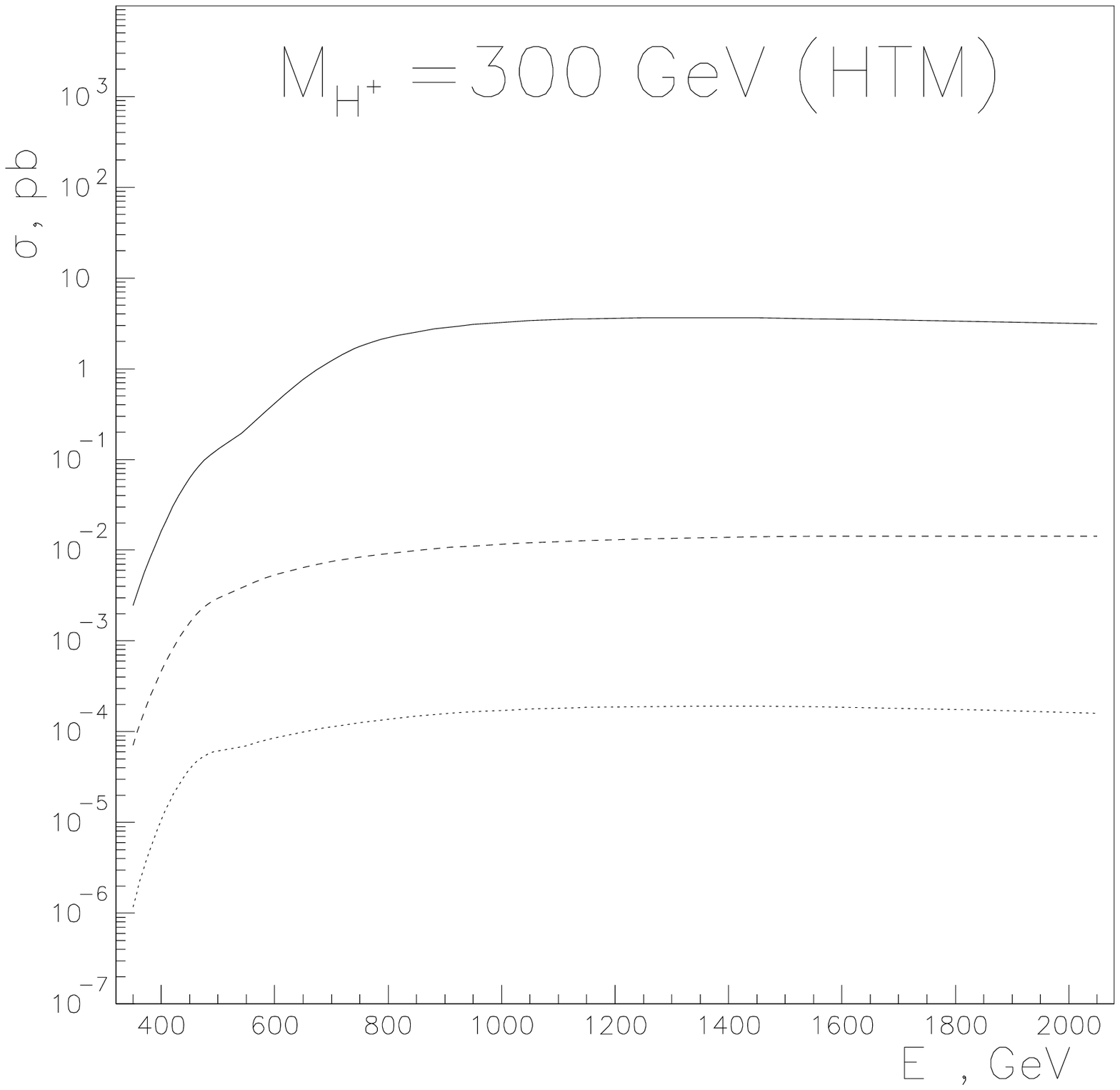}
 \epsfysize=10cm \epsfxsize=7cm
\epsffile{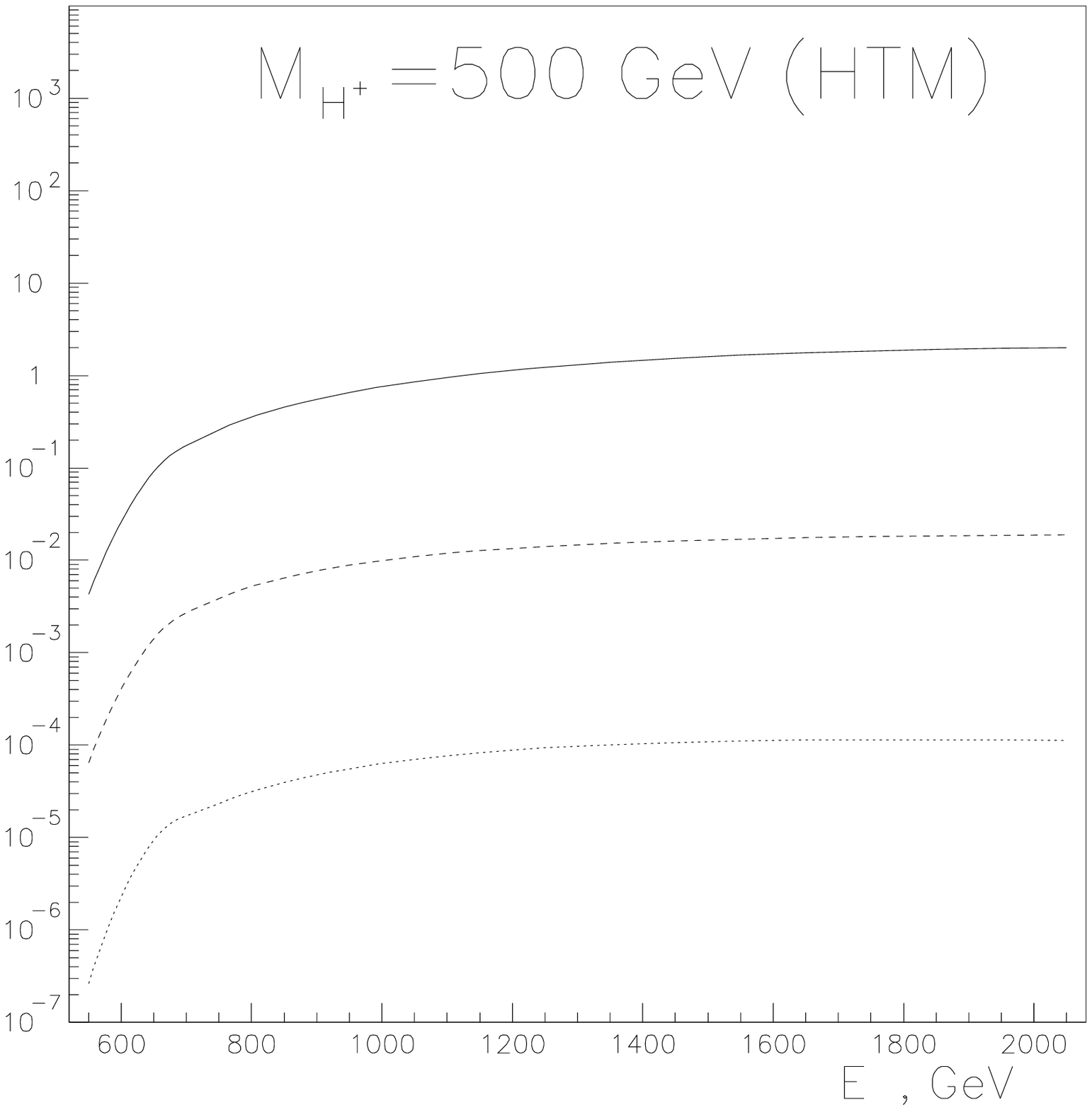}  \\
\hspace*{2.7cm} (a) \hspace*{6.7cm} (b) \\
\caption{The cross section for the $e^+e^-\rightarrow e^-\nu_eH^+$
as a function of center of mass energy for (a) $M_{H^+}=300$ GeV and 
(b) $M_{H^+}=500$ GeV in the HTM.
The triplet Yukawa couplings are $h_{ll}=0$ (dotted),
 $h_{ll}=0.1$
(dashed) and $h_{ll}=1$ (solid), $l=e,\,\mu,\,\tau$.}
\label{triecs}
\end{figure}

Feynman diagrams for the reaction $e^+e^- \rightarrow 
e^- \nu_e H^+$ in the HTM  are presented
 in Fig. \ref{ediag}.
The cross section of this process is plotted in Fig. \ref{triecs}
as a function of centre-of-mass energy $\sqrt{s}$ for the triplet Yukawa
couplings $h_{ll}=0,\,0.1,\,1$ ($l=e,\,\mu,\,\tau$) and $M_{H^+}=300$
GeV and $500$ GeV.
As can be seen, for large Yukawa couplings $h_{ll}$  cross sections  at a few picobarn-level are possible. 
In the limit where all the triplet Yukawa couplings vanish only two 
Feynman amplitudes survive,  which both contain a $WZH$ vertex and are thus suppressed by the factor
$\sin{\theta_H}$.
Therefore the corresponding cross sections, plotted by dotted
lines in Fig. \ref{triecs}, are relatively small.

The cross section for the process $e^+e^-\rightarrow e^-\nu_e H^+$ is in the
2HDM  always very small. 
The reason is that the $H^+ e^- \nu_e$ Yukawa
 coupling is in this model
suppressed by the factor $m_e/M_W$, as can be seen from
 the expression (\ref{coupl}). 
Clearly, this process allows for separation between the two models: 
if one measures a large cross sections for this reaction, it could be a signal 
of the existence of triplet Higgs representations with a large Yukawa coupling
$h_{ee}$. Of course, it is quite naturally to expect
the hierarchy in Yukawa couplings $h_{ee} <  h_{\mu \mu} < 
h_{\tau \tau} $. In this case, when $h_{ee} $ is too small
to observe the above mentioned reaction, the study
of $\mu^+ \mu^- \rightarrow \mu^- \nu{\mu} H^+ $
may be quite useful. For this reaction the results
depicted in Fig. 4 remain unchanged if $h_{ee} =0$.

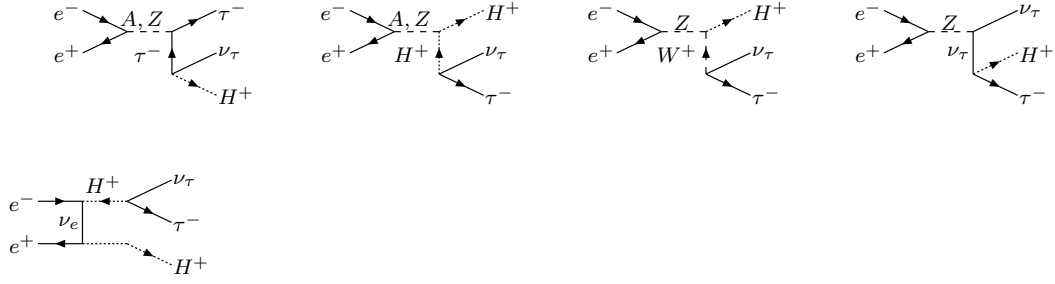
\begin{figure}[t]
{
\unitlength=0.8 pt
\SetScale{0.8}
\SetWidth{0.56}      
\scriptsize    
{} \qquad\allowbreak
\noindent
\begin{picture}(95,79)(0,0)
\Text(15.0,70.0)[r]{$e^-$}
\ArrowLine(16.0,70.0)(37.0,60.0) 
\Text(15.0,50.0)[r]{$e^+$}
\ArrowLine(37.0,60.0)(16.0,50.0) 
\Text(44.0,61.0)[b]{$A,Z$}
\DashLine(37.0,60.0)(58.0,60.0){3.0} 
\Text(80.0,70.0)[l]{$\tau^-$}
\ArrowLine(58.0,60.0)(79.0,70.0) 
\Text(54.0,50.0)[r]{$\tau^-$}
\ArrowLine(58.0,40.0)(58.0,60.0) 
\Text(80.0,50.0)[l]{$\nu_\tau$}
\Line(58.0,40.0)(79.0,50.0) 
\Text(80.0,30.0)[l]{$H^+$}
\DashArrowLine(58.0,40.0)(79.0,30.0){1.0} 
\end{picture} \ 
{} \qquad\allowbreak
\begin{picture}(95,79)(0,0)
\Text(15.0,70.0)[r]{$e^-$}
\ArrowLine(16.0,70.0)(37.0,60.0) 
\Text(15.0,50.0)[r]{$e^+$}
\ArrowLine(37.0,60.0)(16.0,50.0) 
\Text(44.0,61.0)[b]{$A,Z$}
\DashLine(37.0,60.0)(58.0,60.0){3.0} 
\Text(80.0,70.0)[l]{$H^+$}
\DashArrowLine(58.0,60.0)(79.0,70.0){1.0} 
\Text(54.0,50.0)[r]{$H^+$}
\DashArrowLine(58.0,40.0)(58.0,60.0){1.0} 
\Text(80.0,50.0)[l]{$\nu_\tau$}
\Line(58.0,40.0)(79.0,50.0) 
\Text(80.0,30.0)[l]{$\tau^-$}
\ArrowLine(58.0,40.0)(79.0,30.0) 
\end{picture} \ 
{} \qquad\allowbreak
\begin{picture}(95,79)(0,0)
\Text(15.0,70.0)[r]{$e^-$}
\ArrowLine(16.0,70.0)(37.0,60.0) 
\Text(15.0,50.0)[r]{$e^+$}
\ArrowLine(37.0,60.0)(16.0,50.0) 
\Text(47.0,61.0)[b]{$Z$}
\DashLine(37.0,60.0)(58.0,60.0){3.0} 
\Text(80.0,70.0)[l]{$H^+$}
\DashArrowLine(58.0,60.0)(79.0,70.0){1.0} 
\Text(54.0,50.0)[r]{$W^+$}
\DashArrowLine(58.0,40.0)(58.0,60.0){3.0} 
\Text(80.0,50.0)[l]{$\nu_\tau$}
\Line(58.0,40.0)(79.0,50.0) 
\Text(80.0,30.0)[l]{$\tau^-$}
\ArrowLine(58.0,40.0)(79.0,30.0) 
\end{picture} \ 
{} \qquad\allowbreak
\begin{picture}(95,79)(0,0)
\Text(15.0,70.0)[r]{$e^-$}
\ArrowLine(16.0,70.0)(37.0,60.0) 
\Text(15.0,50.0)[r]{$e^+$}
\ArrowLine(37.0,60.0)(16.0,50.0) 
\Text(47.0,61.0)[b]{$Z$}
\DashLine(37.0,60.0)(58.0,60.0){3.0} 
\Text(80.0,70.0)[l]{$\nu_\tau$}
\Line(58.0,60.0)(79.0,70.0) 
\Text(57.0,50.0)[r]{$\nu_\tau$}
\Line(58.0,60.0)(58.0,40.0) 
\Text(80.0,50.0)[l]{$H^+$}
\DashArrowLine(58.0,40.0)(79.0,50.0){1.0} 
\Text(80.0,30.0)[l]{$\tau^-$}
\ArrowLine(58.0,40.0)(79.0,30.0) 
\end{picture} \ 
{} \qquad\allowbreak
\begin{picture}(95,79)(0,0)
\Text(15.0,60.0)[r]{$e^-$}
\ArrowLine(16.0,60.0)(37.0,60.0) 
\Text(47.0,64.0)[b]{$H^+$}
\DashArrowLine(58.0,60.0)(37.0,60.0){1.0} 
\Text(80.0,70.0)[l]{$\nu_\tau$}
\Line(58.0,60.0)(79.0,70.0) 
\Text(80.0,50.0)[l]{$\tau^-$}
\ArrowLine(58.0,60.0)(79.0,50.0) 
\Text(36.0,50.0)[r]{$\nu_e$}
\Line(37.0,60.0)(37.0,40.0) 
\Text(15.0,40.0)[r]{$e^+$}
\ArrowLine(37.0,40.0)(16.0,40.0) 
\DashLine(37.0,40.0)(58.0,40.0){1.0}
\Text(80.0,30.0)[l]{$H^+$}
\DashArrowLine(58.0,40.0)(79.0,30.0){1.0} 
\end{picture} \ 
}
\caption{Feynman diagrams for the process 
$e^+e^- \rightarrow \tau^-\nu_{\tau} H^+$ in the triplet model.}\label{fgs1}
\end{figure}

\begin{figure}[t]
\vspace*{-1.1cm} \hspace*{-0.3cm}
\epsfysize=7cm \epsfxsize=7cm
\epsffile{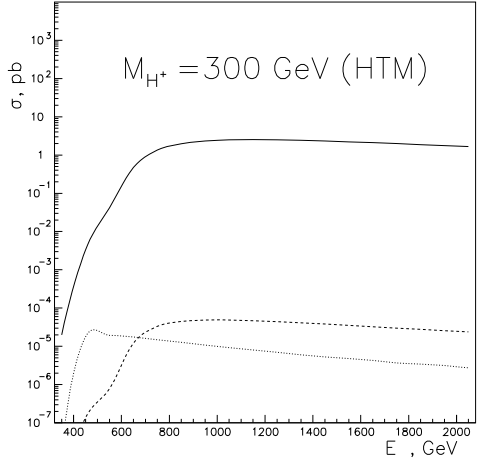}
 \epsfysize=7cm \epsfxsize=7cm
\epsffile{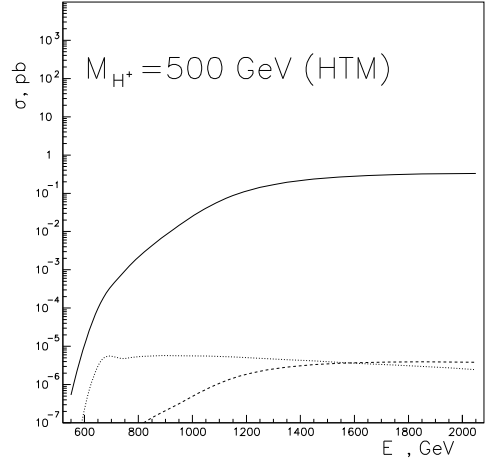} \\
\hspace*{2.7cm} (a) \hspace*{6.7cm} (b) \\
 \hspace*{0.2cm}
 \epsfysize=7cm \epsfxsize=7cm 
\epsffile{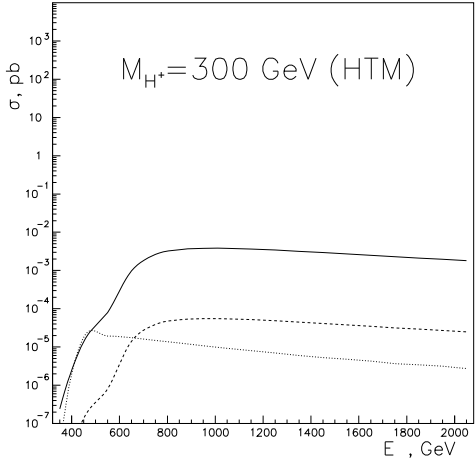} 
\epsfysize=7cm \epsfxsize=7cm
\epsffile{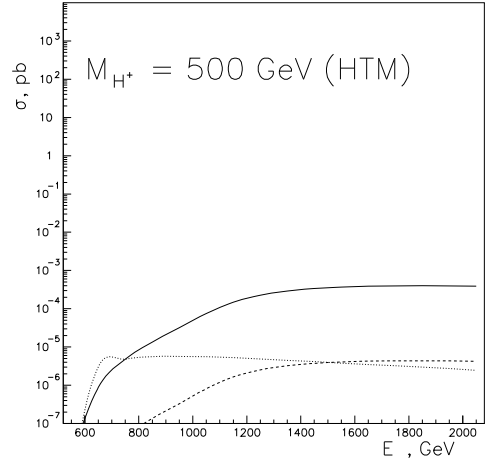} \\
\hspace*{2.7cm} (c) \hspace*{6.7cm} (d) \\
\caption{The cross section for the 
$e^+e^-\rightarrow \tau^-\nu_{\tau} H^+$
as a function of center of mass energy for (a),
 (c) $M_{H^+}=300$ GeV and 
(b), (d) $M_{H^+}=500$ GeV in HTM.
In (a) and (b) the triplet Yukawa couplings are
 taken to be equal for all
three generations: $h_{ll}=1$ (solid), $h_{ll}=0.1$
(dashed) and $h_{ll}=0$ (dotted).
In (c) and (d) only the third generation triplet
 Yukawa coupling is 
nonzero: $h_{\tau \tau}=1$ (solid), $h_{\tau \tau}=0.1$
(dashed) and $h_{\tau \tau}=0$ (dotted).}\label{tritaucs}
\end{figure}

 Feynman diagrams of the process (\ref{process}) 
 with $l=\tau$   are shown in the Fig. \ref{fgs1}
in the HTM. All the amplitudes are proportional to the triplet Yukawa couplings
except the one with a virtual $Z$ boson decaying via the $WZH$ coupling.   
 The total cross sections of 
this process are plotted in  Fig. \ref{tritaucs} (a)-(d) as a function of centre-of-mass energy $\sqrt{s}$. 
In Fig. \ref{tritaucs} (a) and  (b) all the Yukawa couplings are
taken to be equal whereas in Fig. \ref{tritaucs} (c) and (d) only 
the third generation coupling $h_{\tau \tau}$ differs from
zero. 
In the latter case the last diagram of Fig. \ref{fgs1} does not
contribute, since it is proportional to the first generation triplet
Yukawa coupling $h_{ee}$.
The solid and dashed lines in these plots demonstrate the effect 
of the triplet Yukawa couplings.
For simplicity, the triplet VEV $w$ is set to zero in evaluating these curves. 
The dotted lines show the cross section in the case when all the
triplet Yukawa couplings are set to zero and the triplet VEV has a 
nonzero value $w=15$ GeV. 
In this case there is only one diagram which contributes 
to the cross section, the one involving an s-channel $Z$ and a
 $WZH$ vertex. The $\sin{\theta_H}$ suppression
  of the vertex makes the cross section negligible. 

Comparing the Fig. \ref{tritaucs} (a),(c) and \ref{tritaucs}
(b),(d) one can see the large effect of the t-channel neutrino exchange 
diagram that involves the coupling $h_{ee}$. 
If this coupling is $h_{ee}\simeq 1$, the 
cross section is about three orders of magnitude larger than the cross 
section for $h_{ee}=0$. 

For the case of nonvanishing Yukawa couplings,  the 
cross section can be in the region 
$\sqrt{s}>2 M_{H^\pm}$ well approximated with the real $H^\pm$ pair production 
cross section multiplied by the branching ratio of  the decay 
$H^+ \rightarrow \tau \nu$. 
In the region $ \sqrt{s}<2 M_{H^\pm}$ the $H^\pm$-pair production is
not kinematically allowed. 
The single $H^\pm$ production may in this case still be allowed. 
This energy range is therefore of particular interest. 
As seen form the Fig.  \ref{tritaucs} (a) the cross section in this region 
may reach a few hundred femtobarn level for favourable model parameter 
values. 

\begin{figure}[t]
\vspace*{-1.1cm} \hspace*{-0.3cm}
\epsfysize=7cm \epsfxsize=7cm
\epsffile{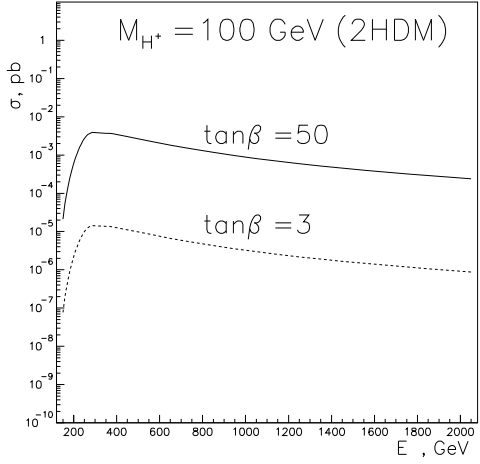}
 \epsfysize=7cm \epsfxsize=7cm
\epsffile{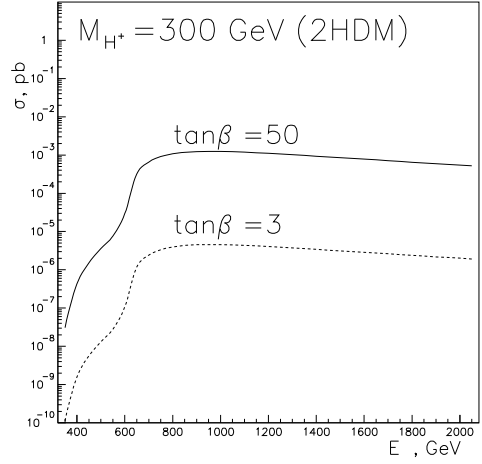} \\
\hspace*{3.2cm} (a) \hspace*{6.3cm} (b) \\
\epsfysize=7cm \epsfxsize=7cm
\hspace*{3.5cm} \epsffile{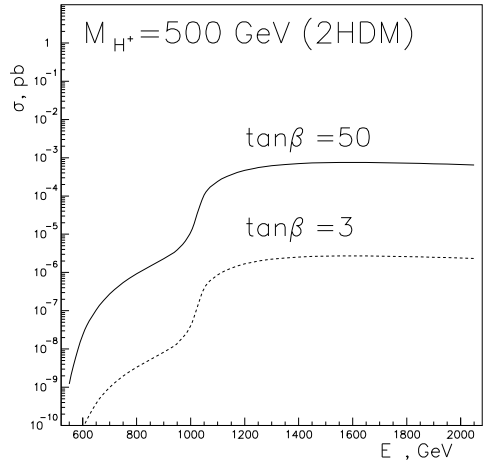} \\
\hspace*{6.7cm} (c)
\caption{The cross section for the $e^+e^-\rightarrow \tau^-\nu_\tau H^+$
as a function of center of mass energy for 
$M_{H^+}=300$ GeV (a) and 
$M_{H^+}=500$ GeV (b) in 2HDM.
The solid line corresponds to
 $\tan\beta=50$ and dashed line
corresponds to $\tan\beta=3$. }
\vspace*{5mm}
\label{doubtaucs}
\end{figure}

The cross sections for the reaction (\ref{process}) with $l=\tau$
in the 2HDM are depicted in the Fig. \ref{doubtaucs} for 
$\tan \beta=3$ and $50$. 
The cross section is larger for larger $\tan \beta$ and is at most a few 
femtobarns for the choice $\tan \beta=50$. 
For $m_{H^+}=300$ GeV the threshold effects of the $H^+H^-$ pair production
can be be seen around $\sqrt{s}=600$ GeV. 
The cross sections for the production of muon and muon neutrino can be 
obtained from the $\tau \nu_\tau$ cross section simply by multiplying by 
$(m_\mu/m_\tau)^2$. 

\begin{figure}[t]
\vspace*{-1.1cm} \hspace*{-0.3cm}
 \epsfysize=7cm \epsfxsize=7cm 
\epsffile{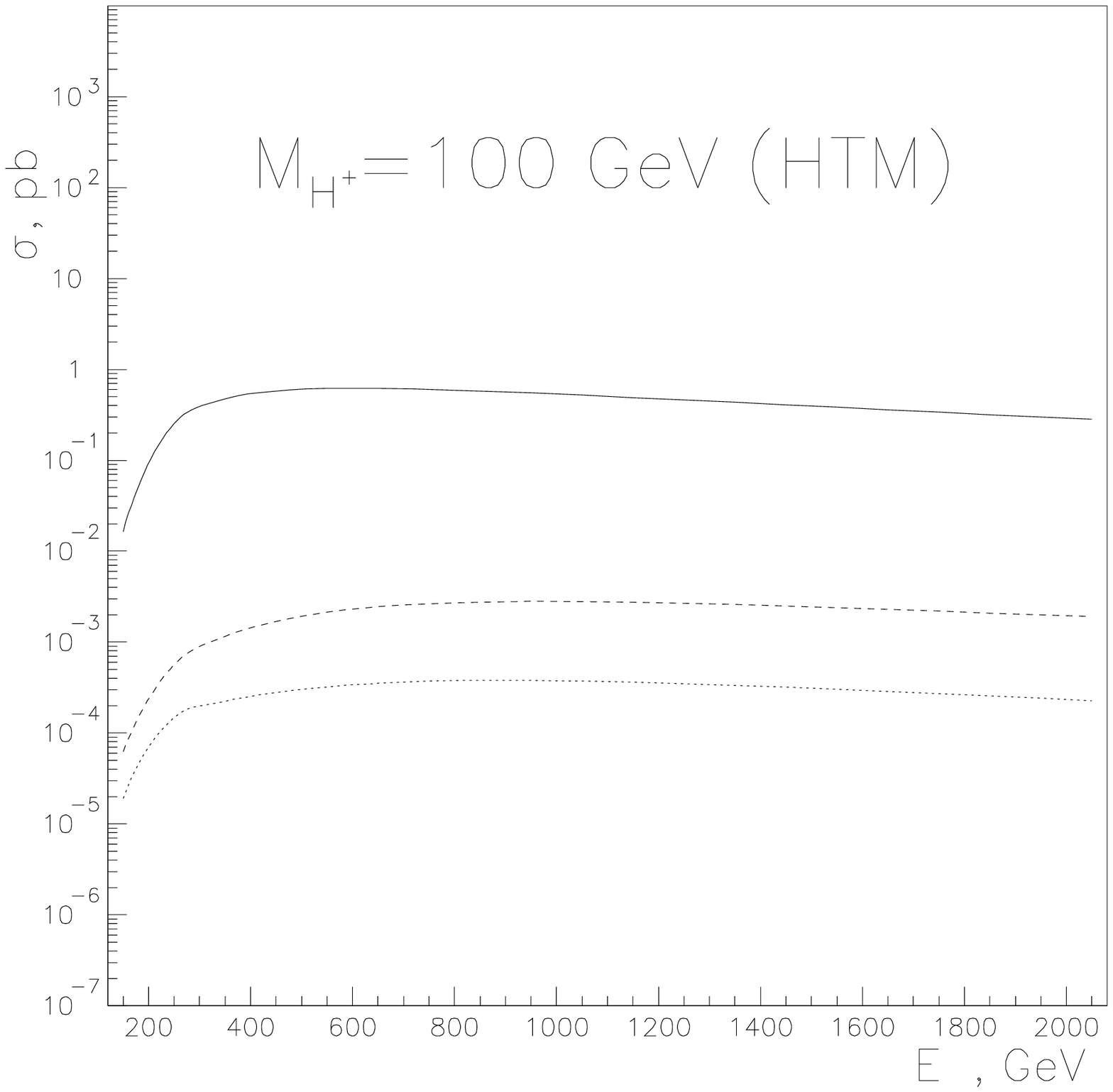} 
\epsfysize=7cm \epsfxsize=7cm
\epsffile{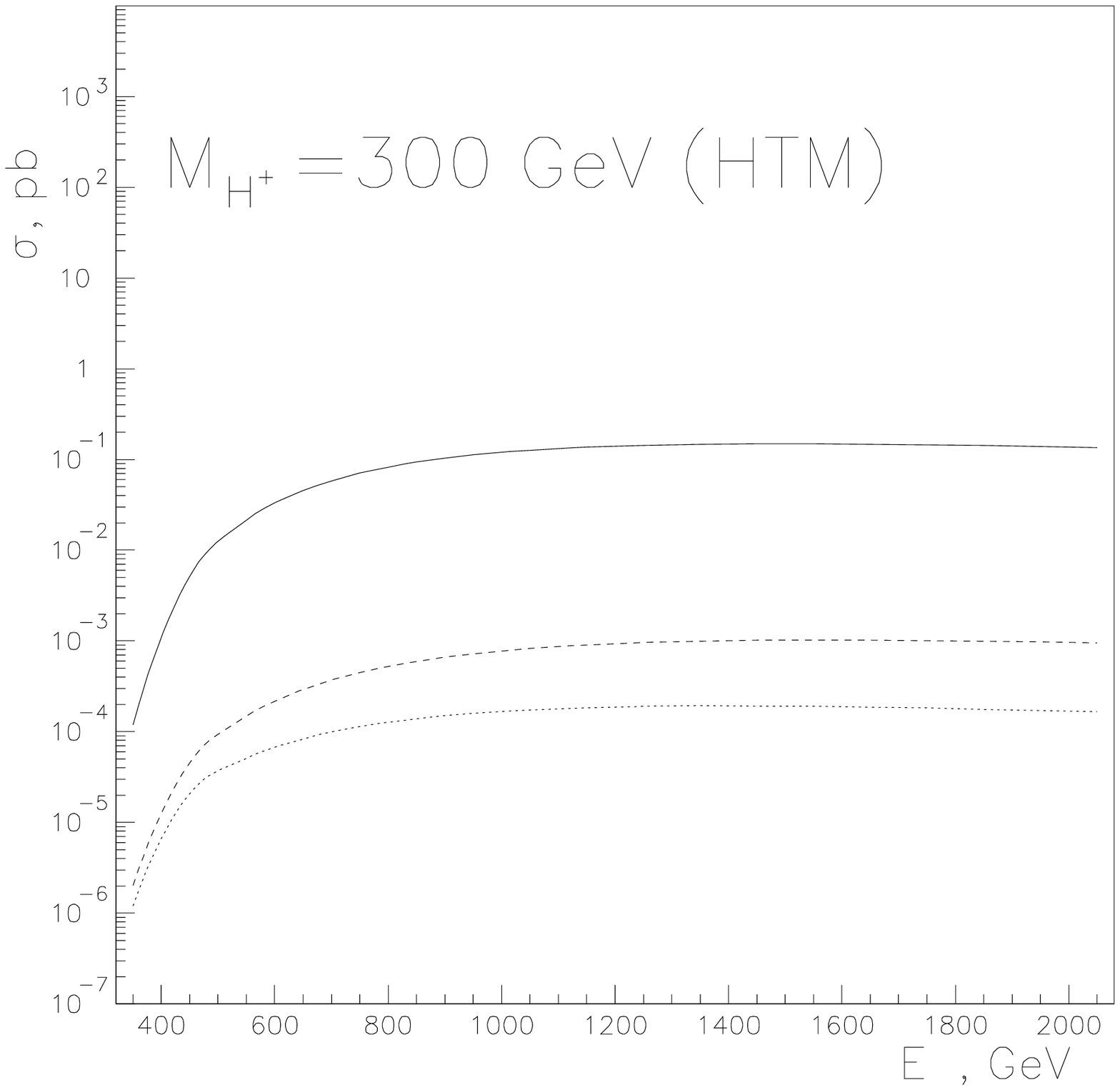} \\
\hspace*{2.7cm} (a) \hspace*{6.7cm} (b) \\
\hspace*{3.2cm}
\epsfysize=7cm \epsfxsize=7cm
\epsffile{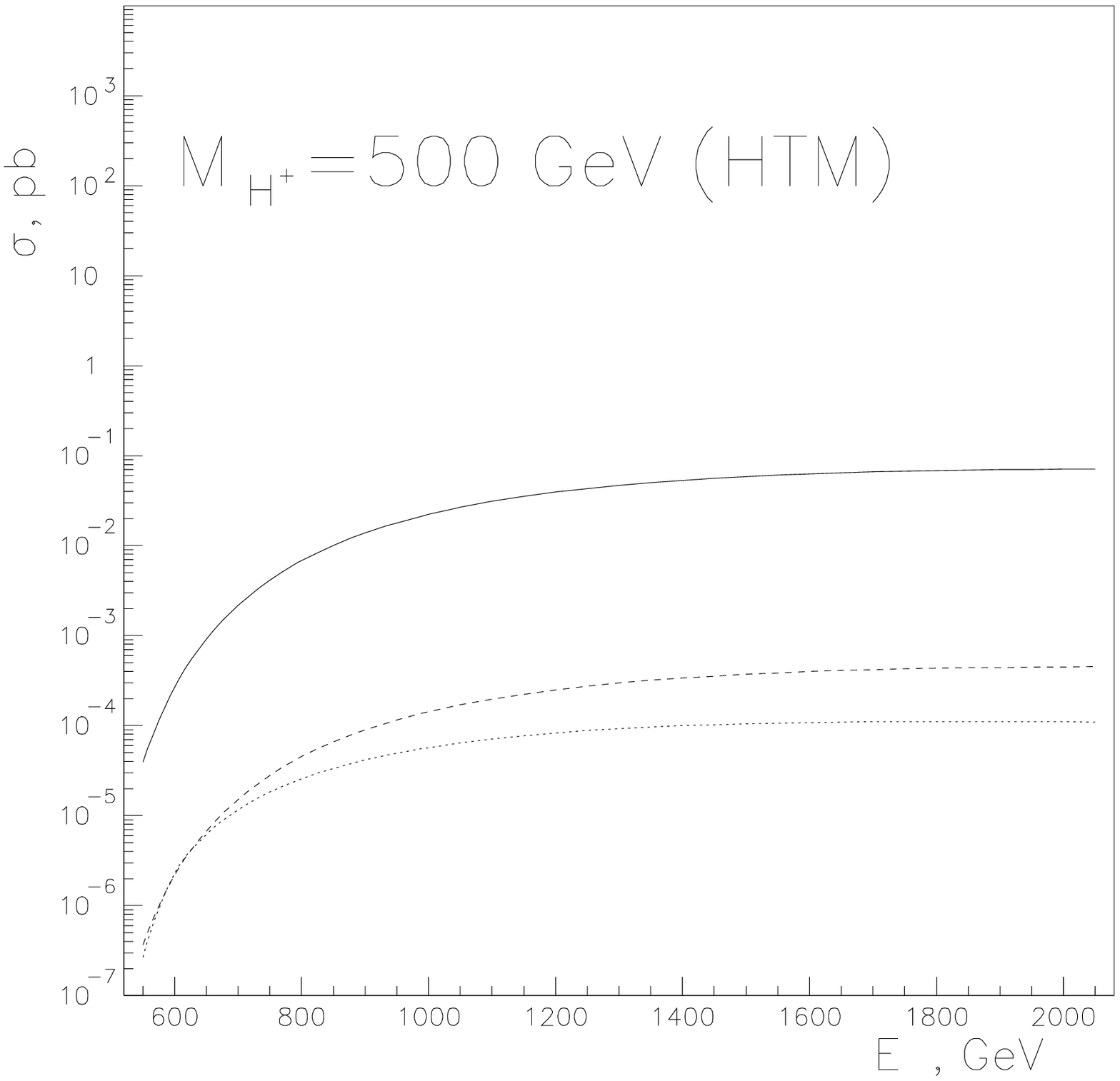} \\
\hspace*{6.4cm} (c)
\caption{The cross section for the $\mu^+e^-\rightarrow \mu^+\nu_e H^-$
as a function of center of mass energy for (a) $M_{H^+}=300$ GeV  and 
(b) $M_{H^+}=500$ GeV in HTM.
The triplet Yukawa couplings are $h_{ll}=1$ (solid),
 $h_{ll}=0.1$
(dashed) and $h_{ll}=0$ (dotted).}
\label{tricsmu}
\end{figure}
 
 We have also considered  the production of a single $H^\pm$ in the reaction 
\be
\mu^+ e^- \rightarrow \mu^+ \nu_e H^-
\ee
in the $e^- \mu^+$  collider.
For this initial state, the pair production of real $H^\pm$'s is not possible. 
The cross sections for this process in the HTM are depicted
in the Fig. \ref{tricsmu} (a) and (b).
The cross sections are at most a few hundred
 femtobarns for large
Yukawa couplings. 

\begin{figure}[t]
\vspace*{-1.1cm} \hspace*{-0.3cm}
 \epsfysize=10cm \epsfxsize=7cm 
\epsffile{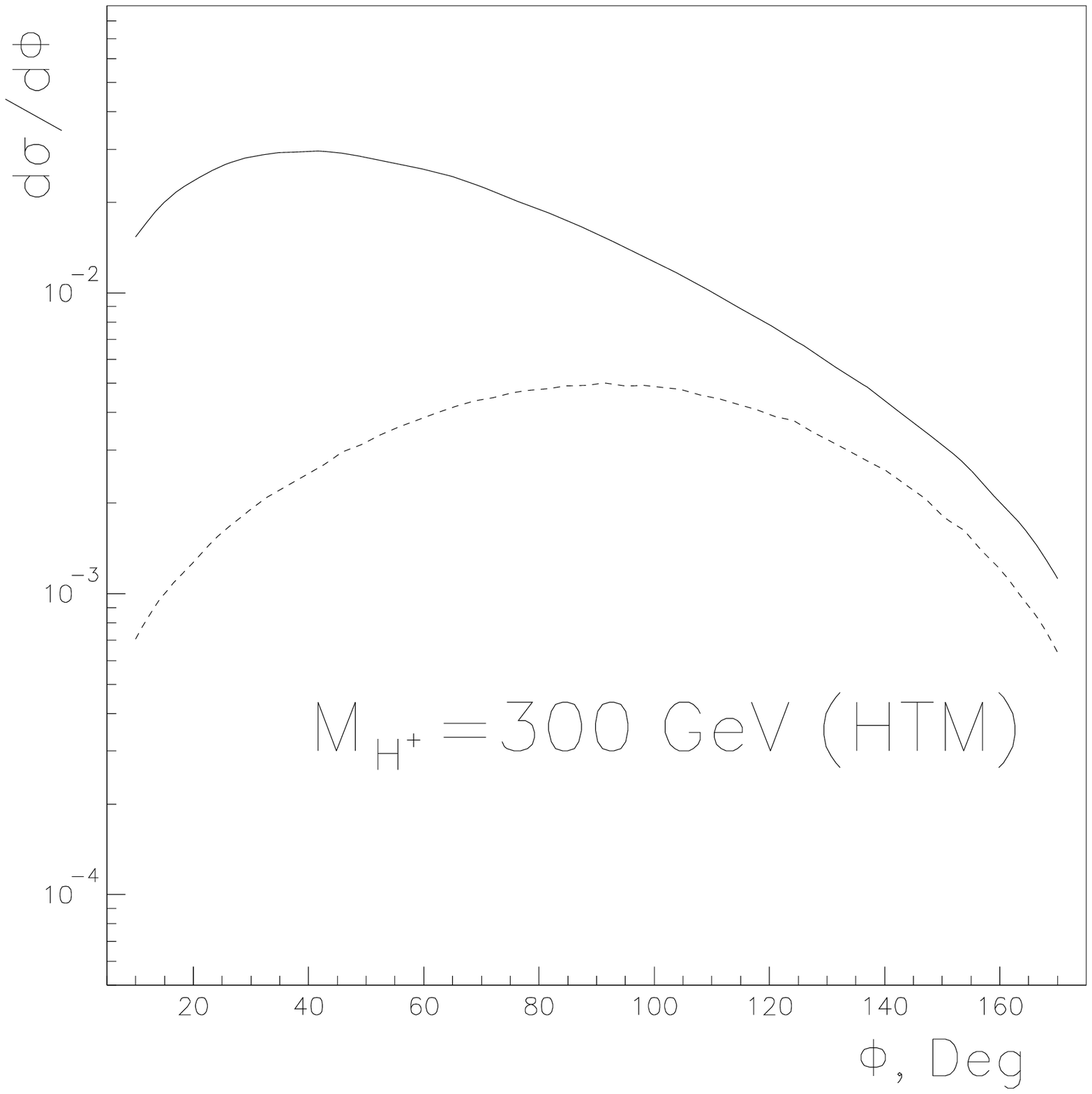} 
\epsfysize=10cm \epsfxsize=7cm
\epsffile{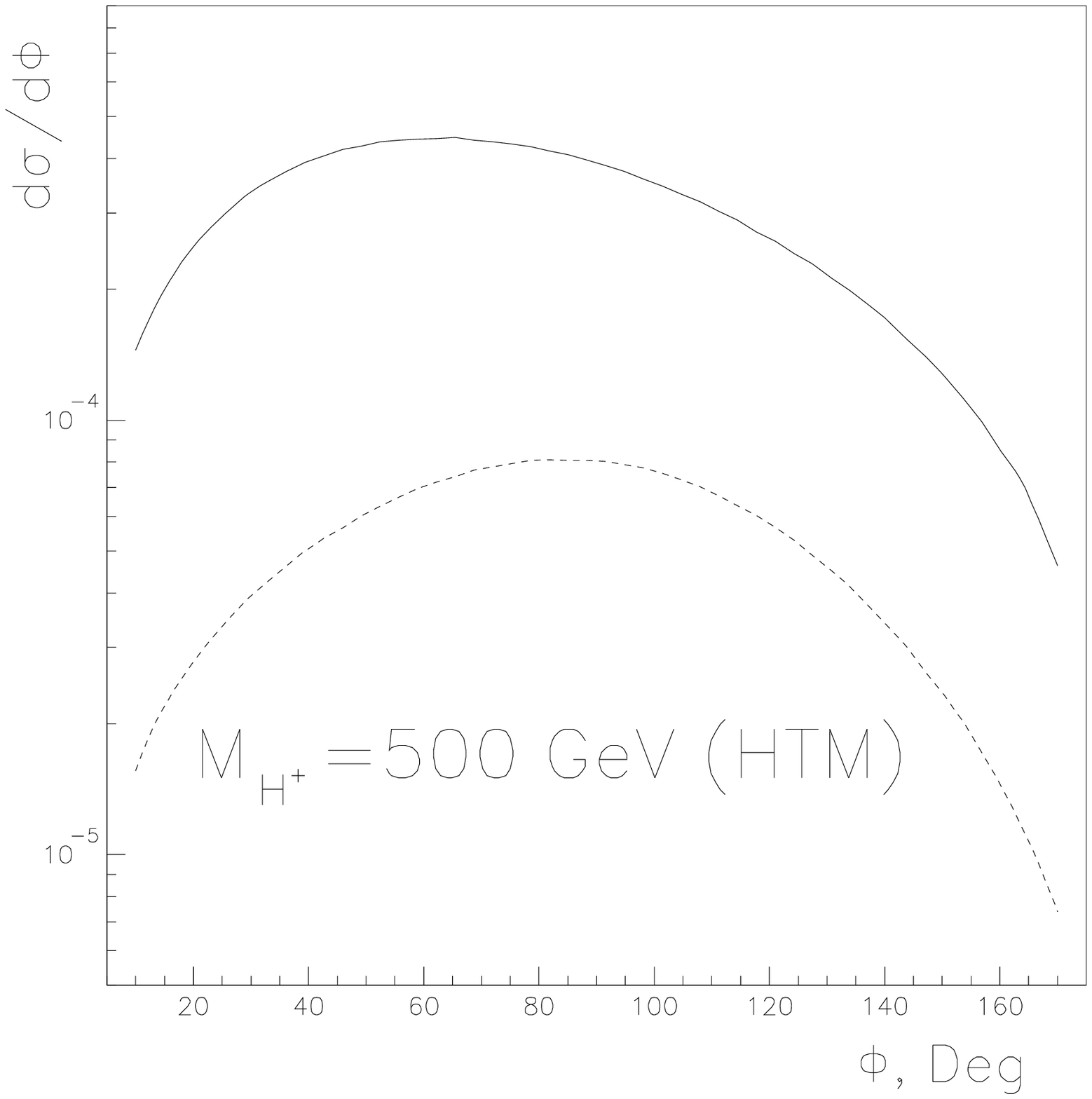} \\
\caption{Angular distributions of the final state $\tau$'s for the
$e^+e^-\rightarrow\tau^-\nu_\tau H^+$ in the triplet Higgs model for
(a) $M_{H^+}=300$ GeV  and (b) $M_{H^+}=500$ GeV.  
Solid lines correspond to $h_{ll}=1$ and dashed lines to
$h_{ll}=0.1$.}\label{tridiff}
\end{figure}

It is in principle possible to make distinction between the HTM and 2HDM by looking at the angular distribution of the outgoing lepton. 
In the 2HDM, the distribution is symmetric and resembles the $\sin^2 \theta$-form of the 
pair production differential cross section, whereas in the HTM
the peak of the distribution is shifted forward.
 The shift is significant only if the 
first generation Majorana Yukawa coupling is large, since in this case there are large
t-channel neutrino exchange diagrams which contribute to the process altering 
the form of the distribution. 
This is illustrated in the Fig. \ref{tridiff}, which shows the 
angular distributions of the charged lepton in the HTM. 
Clearly a deviation from 
the $\sin^2 \theta$ distribution, after the SM background has been removed, would be a signal 
of triplet higgses. The magnitude of this effect, as discussed, depends on the model parameters.


Let us now discuss the SM background to the process (\ref{process}). 
The main background comes from the $W$-pair production $e^+e^- \rightarrow W^-W^+$ 
with subsequent decays $W^\pm \rightarrow \tau^\pm \nu$. The cross section for the 
W-pair production is about $7.5$ pb at $\sqrt{s}=500$ GeV.   
Multiplying by Br($W \rightarrow \tau \nu)^2 \approx 0.013$ gives the value 98 fb for the  
$e^+e^- \rightarrow \tau^- \tau^+ \nu \nu$ cross section.

In this SM background process the final state 
leptons are preferentially emitted in the direction 
of the beam axis, whereas in the signal process (\ref{process}) the angular distribution
of the leptons vanishes in the beam directions. 
Thus, this background can be substantially reduced by requiring the scattered leptons 
to be away from the direction of the beam.  

There is also background from the pair production of $Z$'s with one $Z$ decaying into 
a pair of charged leptons and the other into  a neutrino pair. This background 
can be reduced by requiring the invariant mass of the charged lepton pair to be different 
from $M_Z$.

\section{Conclusions}
\indent
In conclusion, we have investigated the production of  singly charged Higgs 
boson $H^\pm$ at  future linear collider in the framework of two models, the triplet Higgs model (HTM) and two-Higgs-doublet model (2HDM). The aim has been to find out signatures that would allow one to make distinction between the models.

\indent
The single production of $H^\pm$ in the process $e^+e^- \rightarrow l^- \nu H^+$
can be useful in exploring the scalar sector
if the charged Higgs is so heavy that it cannot be pair produced. In the 
HTM, the cross sections depend on the unknown Majorana Yukawa couplings 
and the mass $M_{H^\pm}$ of the charged Higgs. In the 
supersymmetry-motivated 2HDM, the free parameters
 are $M_{H^\pm}$ and 
$\tan \beta$. The  process $e^-e^- \rightarrow e^- \nu_e H^+$  is the most promising one for
 separating  the two models. In the HTM, the cross section for 
this reaction can be of the order of $ 1$ pb for 
$ 100<M_{H^\pm}<300$ GeV.  In the case when $h_{ee} =0$
the same results may be obtained  for
the reaction $\mu^+ \mu^- \rightarrow \mu^- \nu_{\mu} H^+$
at muonic collider.
The reaction  $e^-e^- \rightarrow \tau^- \nu_{\tau} H^+$
is also well observable and provides the separation
between HTM and 2HDM through angular distribution of final
$\tau$'s and decay modes of $H^+$.

In 2HDM, on the other hand,  the couplings of fermions with the
higgses are always 
proportional to the ratio $(m_f/M_W)$. 
This mass hierarchy between Yukawa couplings 
doesn't have to hold in the HTM and its nonexistence can be used to differentiate 
between the two models,   
   since light fermions can have large couplings with 
the charged Higgs boson.   
In our calculations we have  reduced the number of independent parameters in the 2HDM 
by assuming the MSSM supersymmetric relations among the Higgs potential parameters.
The only free parameters in our calculations are then the charged
Higgs mass $m_{H^\pm}$ 
and $\tan \beta$. 

The signatures of the processes at the detector are determined by the decays 
of the charged Higgs. In the 2HDM, the dominant decay channel for the 
charged Higgs is $H^+ \rightarrow \overline{b} t$
 above the threshold. In the HTM this channel is suppressed by the ratio of triplet and doublet vevs.                               
The dominant decay channels in the HTM are likely to be $H^+ \rightarrow l \nu$
or, when kinematically allowed, $H^+ \rightarrow H^0 W^+$. 
Below the $\bar{b}t$ threshold the dominant decay mode in
the 2HDM is $H^+ \rightarrow c \bar{s}$, while in the HTM
the dominant channel is $H^+ \rightarrow l^+ \nu_l$
provided the appropriate Yukawa coupling
$h_{ll} \ge 0.1$.

 \section*{Acknowledgments}

  One of us   (N.R.)  is  grateful 
  to the Theoretical Physics Division
  at the Department of Physics of Helsinki University
  for warm hospitality.
  It is also a great pleasure to thank Alexander Pukhov
  for helpful instructions for using the CompHEP package.
This work has been  supported also by the Academy of Finland under
the contract
 40677  and project number
 163394, and by RFFI grant 98-02-18137.

\end{document}